\documentclass[a4paper,12pt]{article}
\usepackage[utf8]{inputenc}
\usepackage{amsmath}
\usepackage{amssymb}
\usepackage{cancel}
\usepackage{slashed, color}
\usepackage{graphicx}
\usepackage{caption}
\usepackage{subcaption}
\usepackage{mwe}
\usepackage{multirow}
\title{Short Distance Modification of a Gravitational System and its Optical Analog}
\author{Mir Faizal$^{1,2}$, Qin Zhao$^3$, Chenguang Hou$^3$,  Zaid Zaz$^{4,5}$\\
$^1$ Irving K. Barber School of Arts and Sciences, \\ University of British Columbia - Okanagan, 3333
University Way, \\ Kelowna, British Columbia V1V 1V7, Canada.  \\$^2$ Department of Physics and Astronomy,
University of Lethbridge, \\
Lethbridge,   Alberta, T1K 3M4, Canada. \\ 
$^3$ Department of Physics, National University of Singapore, \\
 2 Science Drive 3, Singapore.\\  $^4$ Theoritical Physics Division,  Department of Physics, \\ National Institute of Technology, \\
Srinagar, Kashmir-190006, India 
\\$^5$ Department of Electronics and Communication Engineering, \\University of Kashmir, 
Srinagar, Kashmir-190006, India}

\date{}
\begin{document}

\maketitle

\begin{abstract}
Motivated by developments in string theory, such as T-duality, it has been proposed that the geometry of spacetime should have an 
intrinsic minimal length associated with it. 
This would modify the short distance behavior of    quantum systems studied on such a geometry, and an  optical analog for such 
a short distance modification of quantum system has also been realized by using  non-paraxial nonlinear optics. 
As general relativity can be viewed  as an effective field theory obtained from string, it is expected that this would 
also modify the short distance behavior of general relativity. Now the Newtonian approximation is a valid 
short distance approximation to general relativity, and Schrodinger-Newton equation can be obtained as a non-relativistic semi-classical 
limit of such a theory, we will analyze the short distance modification of Schrodinger-Newton equation from an intrinsic minimal length in 
the geometry of spacetime.  As an optical   analog of the Schrodinger-Newton equation has been constructed, 
it is possible to optically realize this system. So, this system is important, and we will numerical analyze the solutions for this system. 
It will be observed that the usual Runge-Kutta method cannot be used to analyze this system. 
However, we  will use a propose and use a new  numerical method, which we will call as the 
two step Runge-Kutta method,  for analyzing this system.  
\end{abstract}
 
\section{Introduction}
It is known that general relativity is an effective field theory approximation to some more fundamental theory, such as the string theory. 
So, we would expect some features of this fundamental theory to modify the short distance behavior of the general relativity. 
An interesting feature of string theory is that there is an intrinsic minimal length scale in string theory, and it is not possible 
to define the geometry of spacetime below that length scale    \cite{z2}-\cite{2z}. 
This is because strings have an extended structure, and the 
  fundamental string is the smallest probe  in perturbative string theory. So, it is not possible to probe the geometry 
  of spacetime below the string length scale . 
  Thus, the   string length acts, which is given by $l_s = \alpha'$ as   a minimum measurable length in string theory.
Even thought to analyze   non-perturbative effects in string theory, we also need to consider $D0$-brane, and $D0$-branes are point 
like objects, it can still be argued that there would exist a minimal length in non-perturbative string theory.
In fact, it has been demonstrated that   an intrinsic minimal length $l_{min}$ exists even if $D$-branes scattering is considered 
\cite{s16}-\cite{s18}. 
Here this minimal length $l_{min}$ is related to string length $l_s$ as $l_{min} = l_s g_s^{1/3}$,
where $g_s$ is the string coupling constant  \cite{s18}. 

It can be argued from T-duality that such a minimal length scale would naturally 
exist in string theory. This can be done by analyzing the 
    total energy of the quantized string. Now if we if we consider  one additional
dimension compactification  on a radius $R$, then this total energy would  depend  on the  excitation $n$ and winding number $w$ \cite{s16}. 
    Now   then under T-duality, 
it is possible to interchange the excitation $n$ and the winding number $w$, such that 
$     R \to l_s^2/R, $ and $  n \to w.
$    This implies that it is not possible to    describe  string theory below $l_s$. This is because
    the description of string below $l_s$  is the same as the 
  description above it.  So, using T-duality it can be argued that the geometry of spacetime in string theory has an intrinsic
  length associated with it    \cite{s16}.  It may be noted that 
  the  T-duality has also  been used to analyze  the Green's function for the 
 the center of mass of the string using an effective path integral   \cite{green1}-\cite{green2}. 
It has been observed that there     a minimal length also exists in this Green's function, and it is not possible to probe length scales below 
that minimal length in this formalism \cite{green1}-\cite{green2}. So, a minimal length exists in string theory because of the 
   T-duality. Now as general theory of relativity is a low energy effective field theory approximation 
  of string theory, it can be argued that the short distance behavior of general relativity should be modified such that there 
  is an intrinsic minimal length scale associated with it. It mat be noted that such a minimal length also occurs in other 
  branches of quantum gravity such as the loop quantum gravity \cite{z1}. 
  
  It can be argued using black hole physics, 
  that any theory of quantum gravity should have a minimum length at least of the order of Planck length $l_{PL}$
  associated with it \cite{z4}-\cite{z5}. 
  This is because energy needed to probe a region of spacetime below Planck scale is more than the energy needed to form a mini black hole 
  in that region. However, it is possible for the minimal length scale to be larger than the Planck scale. Even in string theory, 
  the minimal length scale, which is the string length scale, can be  greater than the Planck scale. 
   This is because the   Planck length $l_{PL}$ can be expressed in terms of string length $l_s$  
as $l_{PL} = g_s^{1/4} l_s$, where $g_s$ is the string coupling constant   \cite{s16}.   In fact, it has been argued that the 
minimal length scale can be several orders of magnitude greater than Planck length scale, and this can produce 
universal short distance corrections to all quantum mechanical systems \cite{54}. As this   
  minimal length scale can have low energy consequences, and so it can be phenomenologically fixed using low energy 
  experiments \cite{51}.  In fact, it is possible to use  opto-mechanical systems to measure such 
  short distance modification of the quantum mechanics by the 
existence of a minimum measurable length scale  
\cite{Pikovski:2011zk}. However, it   is also possible  to test the Schrodinger-Newton equation experimentally  using
opto-mechanical systems \cite{Gan:2015cxz}-\cite{Grossardt:2015moa}. The  Schrodinger-Newton equation is a semi-classical 
approximation to quantum gravity, and it can be obtained as   the 
non-relativistic limit of  the Dirac equation and the  Klein–Gordon equation with a
classical Newtonian potential \cite{rela12}. The  Schrodinger-Newton 
equation has been used to describe interesting  properties of 
 gravitational systems at such short scales \cite{equation}-\cite{equation1}. It is thought to be a valid semi-classical approximation 
 as Newtonian gravity has been demonstrated to be a valid short distance approximation to general relativity to
 smallest scales at which general relativity  has been tested, which is about $0.4mm$ \cite{mz4}. 
 
As it is possible to test the Schrodinger-Newton  experimentally  using
opto-mechanical systems \cite{Gan:2015cxz}-\cite{Grossardt:2015moa}, and it is also possible 
to use opto-mechanical systems to test short distance modification to quantum mechanical \cite{Pikovski:2011zk}, it is important 
to analyze the short distance modification to  Schrodinger-Newton equation. 
So, in this paper, we will study such a short distance modification of the Schrodinger-Newton equation, 
and analyze this system numerically. We will demonstrate that the usual 
 Runge-Kutta method  cannot be used to obtain good numerical results due the existence 
 of higher derivative terms. We will then propose a new two-step Runge-Kutta method, 
 which will be demonstrated to be free from errors generated in the usual one-step 
 Runge-Kutta method. Thus, the methods of this paper, can used to analyze other GUP 
 deformed systems. 
 
\section{Schrodinger-Newton Equation and its Optical Analog}

It is known that the general relativity  cannot be quantized using the usual methods of 
quantum field theory, and even though we have various proposals for a quantizing 
general relativity, 
we do not still have a fully consistent quantum theory of quantum gravity. 
This has motivated the study of semi-classical quantum gravity, 
and in this approach,   the gravitational
filed is treated as a background classical field, and the 
  matter fields are  treated quantum mechanically.  Thus, 
  in the semi-classical approximation, a  Einstein tensor $G_{\mu\nu}$ 
  is produced by quantum mechanical energy-momentum tensor for matter fields  $\hat{T}_{\mu\nu}$. 
  So, if $|\psi\rangle $ is the wave function of the matter fields, then we can write 
  the Einstein equation in semi-classical approximation as 
  \begin{equation}
G_{\mu\nu} =  \frac{8\pi G}{c^2} \langle \psi |\hat{T}_{\mu\nu}|\psi \rangle. 
\end{equation}
It is known that till the smallest scales at which gravity has been tested ($0.4mm$), 
the Newtonian gravity is a good approximation 
to general relativity \cite{mz4}. Thus at small distances, 
we expect that the semi-classical 
approximation to be described by a Schrodinger-Newton equation, 
\cite{1newt}-\cite{newt0} 
 \begin{equation}\label{action_normal}
i\hbar \frac{\partial\psi}{\partial t} 
= H \psi = \frac{1}{2 m}\hat{P}^2\psi + m\Phi(\textbf{R}, t)\psi,
\end{equation}
where  $H$ is the Hamiltonian operator of a system,  with     
    $m|\psi(\textbf{R}, t)|^2 
= \rho(\textbf{R}, t)$ as the mass density and $\Phi(\textbf{R}, t)$ as the classical 
Newtonian potential. It may be noted that this potential also satisfies the Poisson equation
 \begin{equation}\label{action_normal}
 \nabla^2\Phi(\textbf{R}, t) = - 4 \pi G m |\psi|^2. 
\end{equation}

It is possible to construct a gravity analog for this system, and use it to analyze its properties. 
The gravity analogue are used for studding various   gravitational effects, and this done  using artificial systems. 
These systems recreate some specific properties 
of  the   gravitational  system, and can be experimentally realized in the    laboratory \cite{Barcelo:2005fc}. 
These  gravity analogs have been used to study   analogous   black  holes, and analyze    the
 analogous Hawking radiation in   in transonic fluid flows \cite{Unruh:1980cg}. 
 Such flows have been realized in various physical systems, such as 
flowing  water  \cite{Weinfurtner:2010nu},  Bose-Einstein  condensates \cite{Lahav:2009wx}-\cite{Steinhauer:2014dra}  and
nonlinear optics \cite{Philbin:2007ji}-  \cite{Belgiorno:2010wn}.
These gravity analogs are linear, but gravity is a very nonlinear theory. However, it is possible to realize the non-linearity 
in gravity analogs by using optical wave packets with  thermal nonlinearity, and this system is 
mathematically equivalent to the Newton–Schrodinger  equation  \cite{mathematically}-\cite{mathematically1}.  
This nonlinear gravity analog that can be constructed in 
  laboratory experiments is based on the 
  evolution of the amplitude $\mathcal{E}$, 
of an optical beam in a thermally focusing medium.
This system is described by the following equation, 
\begin{equation}
 i \frac{\partial \mathcal{E} }{\partial z} +\frac{1}{2k} \nabla^2_2 \mathcal{E}
 + k_0 \Delta n \mathcal{E} =0, 
\end{equation}
where $\nabla_2^2$ is the transverse two-dimensional Laplacian.  
Here we also have 
\begin{equation}
 k = n_b \omega/c = n_b k_0, 
\end{equation}
where $n_b$ is the background refractive. The nonlocal change in refractive-index 
$\Delta n$ can be 
induced by heating  the medium by a beam, and it can be expressed
as 
\begin{equation}
 \nabla^2_2 \Delta n = \alpha \beta \kappa^{-1} |\mathcal{E}|^2, 
\end{equation}
where  $\kappa$ 
is the thermal conductivity, $\beta$ 
is  the   thermo-optic  coefficient, and $\alpha$ 
is the absorption coefficient. This system can be used as a analog for the Newton–Schrodinger  equation. 
The advantage of the optical analog of Newton–Schrodinger  equation is that this system can be realized in laboratory, 
and its properties can be experimentally studied using such an analog. 

\section{Deformed Newton–Schrodinger  Equation}
It is important  to analyze the short distance modification to the Schrodinger-Newton equation. 
This is because both the short distances effects from the Schrodinger-Newton equation \cite{Gan:2015cxz}-\cite{Grossardt:2015moa}, 
and short distance modification of quantum mechanics \cite{Pikovski:2011zk}, can be measured  
   opto-mechanical systems. 
  Such a short distance modification to a quantum system  occurs due to the existence of a minimal length scale in the geometry of spacetime, 
and it can be analyzed using  a generalization of  the 
uncertainty principle to a generalized uncertainty principle (GUP) \cite{1}-\cite{14}.  
This generalization also deforms  the  
Heisenberg algebra \cite{17}-\cite{53}, 
\begin{equation}\label{GUP}
[\hat{X}_i, \hat{P}_j] = i\hbar(\delta_{ij} + \beta\delta_{ij}\hat{P}^2 + 2\beta \hat{P}_i \hat{P}_j).
\end{equation}
This deformation of the  Heisenberg algebra also 
deforms coordinate representation of the 
momentum operator  \cite{18}-\cite{10}.
It has been demonstrated that this deformed  Heisenberg algebra satisfies 
the  Jacobi identity, it is possible to demonstrate \cite{54}-\cite{51}
\begin{equation}
[\hat{X}_i, \hat{X}_j] = 0 = [\hat{P}_i, \hat{P}_j]. 
\end{equation}
We expect this short distance modification of the Heisenberg algebra to reduce to the usual deformation of the 
Heisenberg algebra, at low energies. So, if  $\hat{p}_i$ is the momentum at low energies, then we expect that 
\begin{eqnarray}
 \hat{p}_j = -i\hbar\frac{\partial}{\partial {\hat{x}}_j}, 
\end{eqnarray}
where $\hat{x}_i$ is the coordinate conjugate to $\hat{p}_i$, such that 
\begin{eqnarray}
 [\hat{x}_i, \hat{p}_j] = i\hbar\delta_{ij}.  
\end{eqnarray}
Now we can express the  
that at higher energies $\hat{P}_j$, and the coordinate conjugate to it $\hat{X}_i\psi(\textbf{x})$
in terms of the low energy momentum  and coordinates as  \cite{54}-\cite{51}
\begin{eqnarray}
&& \hat{X}_i\psi(\textbf{x}) = \hat{x}_i \psi(\textbf{x}), \\
&& \hat{P}_i\psi(\textbf{x}) = \hat{p}_i(1+\beta \hat{p}^2)\psi(\textbf{x})
\end{eqnarray}
This is because by using this representation,  the first order in $\beta$, (\ref{GUP}) is satisfied.
Here we neglect terms of order $\beta^2$ and higher. It is interesting to note that this deformation of the 
momentum operator has also
been motivated by a non-anticommutative
deformation of a supersymmetric field theory 
\cite{field}. It is interesting  to note that 
a GUP like deformation of the analog Newton-Schrodinger equation can also be performed, as it has been observed that 
a GUP like deformation of optical propagation of focused laser beams occurs in the  non-paraxial nonlinear optics 
\cite{Conti:2014jsa}-\cite{Braidotti:2016ido}.
This is done by analyzing the  propagation of light  beyond the paraxial approximation, and then expanding to the first order. 
So, to the first order a simple non-paraxial system 
deforms to $ - (2 k_0)^{-1} \partial_x^2 \mathcal{A}$ to 
$ - (2 k_0)^{-1} \partial_x^2 \mathcal{A} +
(8 k_0)^3 \partial_x^4 \mathcal{A} $, where $\mathcal{A} = 
\mathcal{E} e^{-k_0 z}$.
This is the same 
deformation produced by GUP, 
if  $\beta = 3 \lambda^2/ 8 h^2 $ \cite{Braidotti:2016ido}. 

Substituting $\hat{X}_i, \hat{P}_i$ by $x_i, p_i$, we can write the GUP deformed 
Schrodinger-Newton equation as (to the leading order in $\beta$) \cite{gravity0}-\cite{garvitya}
\begin{equation}\label{action_gup}
i\hbar \frac{\partial\psi}{\partial t} = H \psi = \frac{\hat{p}^2}{2 m}
\psi + \frac{ \beta}{ m} \hat{p}^4\psi + \Phi(\textbf{r}, t)\psi.
\end{equation}
It may be noted that such a GUP deformed Schrodinger-Newton equation has been used to motivate various studies 
 \cite{gravity0}-\cite{garvitya}. 
We will analyze the   spherical symmetric solutions to this equation, as 
those solution have important physical applications. 
Thus, we can write (\ref{action_gup}) as 
\begin{eqnarray}
-\frac{\hbar^2}{2mr^2}\frac{d}{dr}r^2\frac{d\psi}{dr} +\frac{\hbar^4 \beta}{ m} \frac{d}{r^2 dr}r^2\frac{d}{dr} 
(\frac{d}{r^2 dr}r^2\frac{d\psi}{dr}) + \Phi\psi =  E\psi,\\
\frac{d}{r^2dr}r^2\frac{d\Phi}{dr} = 4\pi Gm^2\psi^2.
\end{eqnarray}
We assume that $\psi$ and $\Phi$ approach zero as
$|\textbf{r}|\rightarrow \infty$ and are regular near the origin. Here we have obtained the short distance modification to the 
semi-classical gravity with an intrinsic minimal length in the geometry. Now we need to understand such a solution, and we shall numerically 
analyze such a soltion.

\section{Usual Runge-Kutta Method }
In this section, we will analyze this system using the usual Runge-Kutta method. 
We will observe that there are several problems with the application of the usual 
Runge-Kutta method to this system. To analyze this system using the usual 
Runge-Kutta method,  we redefine 
\begin{align}
\psi = \zeta  S,  && 
 E - \Phi = \xi V,  
\end{align}
where 
\begin{align}
\zeta = \left(\frac{\hbar^2}{8\pi Gm^3}\right)^{1/2}, && \xi =\frac{\hbar^2}{2m},
\end{align}
and substitute them into the GUP deformed Schr\"odinger-Newton equation, 
\begin{align}\label{dimensionless_emo}
\frac{d^2}{r dr^2}(rS) - 2\hbar^2 \beta \frac{d^4}{r dr^4}(rS)  = - VS,\\
\frac{d}{r^2dr}r^2\frac{d V}{dr} = - S^2.
\end{align}
$S$ and $V$ have dimension (length)$^{-2}$, and the formulas 
\eqref{dimensionless_emo} is invariant under a scale transformation:
\begin{equation}\label{rescaling}
(S, V, \beta, r) \rightarrow (\lambda^2 S, \lambda^2 V, \lambda^{-2} \beta, \lambda^{-1}r).
\end{equation}

It may be noted that because of the rescaling freedom \eqref{rescaling}, one 
can fix $V_0 =1$ and allow $S_0$ to vary, where $V_0$ and $S_0$
are the values of $V$ and $S$ at $r=0$, respectively. Now,  
we can directly attempt to find the numerical solutions by using
fourth-order Runge-Kutta NAG routine. Then, first, we need to
determine the initial values for the different orders of the
derivatives of $S$ and $V$. With bounded derivatives at $r=0$, we assuming that 
\begin{align}
S=S_0 + C_i r^i, && V = 1+ D_i r^i. 
\end{align}
According to \eqref{dimensionless_emo} and comparing the power of $r$, we have 
\begin{align}\label{coefficient}
\begin{aligned}
  & 2C_1 - 12\hbar^2\beta C_3 = 0,\\
&  6C_2 - 120\hbar^2\beta C_4 = -S_0, \\
 & 12 C_3  - 720\hbar^2\beta C_5 = -S_0 D_1 - C_1,\\
 & 20C_4 - 1680\hbar^2\beta C_6 = -S_0 D_2 - C_2,\\
 & \cdots\\
 & 2D_1 = 0, \\
 & 6D_2 = -S_0^2,\\
 & 12 D_3 = -2S_0 C_1, \\
 & 20 D_4 = -2S_0 C_2,\\
 & \cdots 
\end{aligned}
\end{align}
We can set $C_{2i+1}$ and $D_{2i+1}$ to be zero.
Therefore, the first order derivative $S_0^{(1)}$ is zero, 
so is $V_0^{(1)}$. Since the highest derivative of $S$ in \eqref{dimensionless_emo} 
is the fourth order, we still need to initialize the second and the third 
order derivative, $S_0^{(2)}$, $S_0^{(3)}$. It is obvious that $S_0^{(3)}=0$. 
However, since $C_{2i}$ are not independent, we can not exactly solve the equations 
\eqref{coefficient}. If we assume that $\hbar^2\beta$ very small, approximately,
we have $S_0^{(2)} = 2 C_2 = -\frac{S_0}{3}$.

With this set-up, we solve the formulas \eqref{dimensionless_emo}. However,
when we consider $\beta > 0$, due to positive iterations (we means that $S^{(4)}$
equals to  $S^{(2)}$ multiplying a positive coefficient and plus something) which can easily lead to a divergence, 
we can not obtain stable and smooth solutions. Therefore, for test, we consider $\beta < 0$. 

\begin{figure*}
        \centering
        \begin{subfigure}[b]{0.475\textwidth}
            \centering
            \includegraphics[scale=0.3]{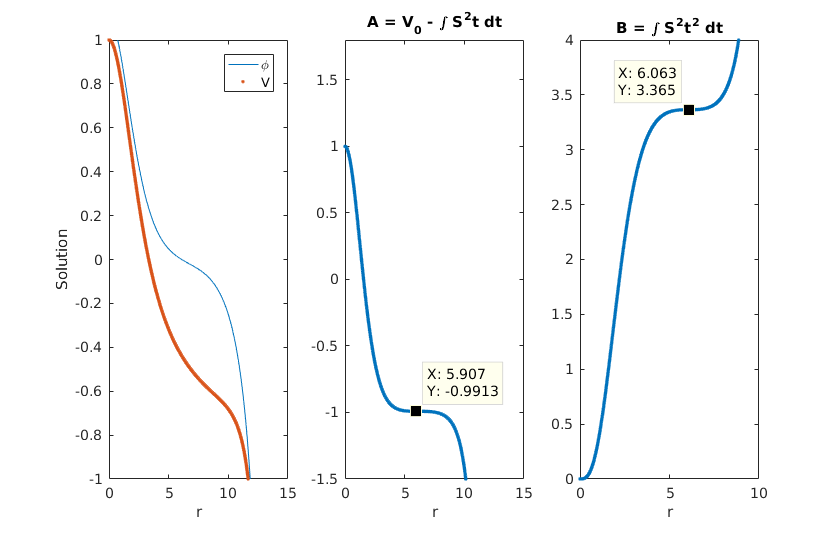}
            \caption[]%
            {{\scriptsize $S_0 =1.0894125784434689486 $, $A = -0.1.065 $, $B = 3.615$}}    
            \label{fig:mean and std of net14}
        \end{subfigure}
        \hfill
        \begin{subfigure}[b]{0.475\textwidth}  
            \centering 
            \includegraphics[scale=0.3]{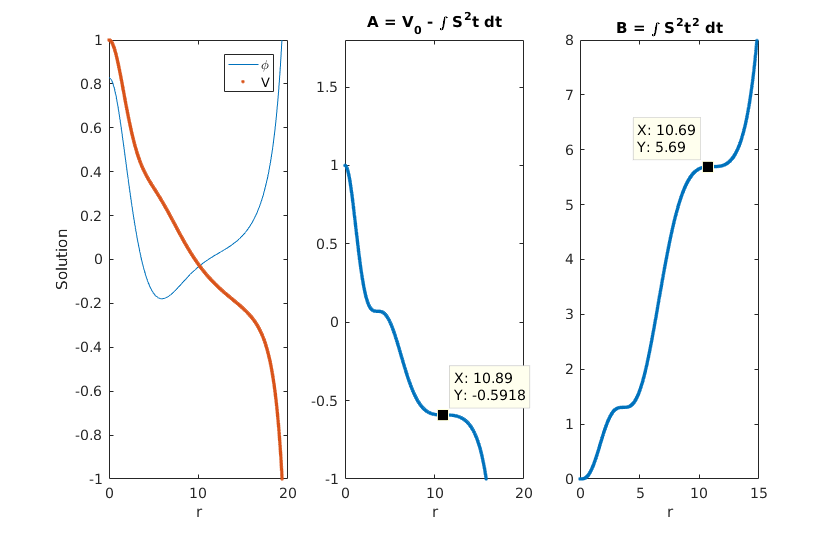}
            \caption[]%
            {{\scriptsize $S_0 =0.82666388605282536961 $, $A = -0.5918 $, $B = 5.69$}}    
            \label{fig:mean and std of net24}
        \end{subfigure}
        \vskip\baselineskip
        \begin{subfigure}[b]{0.475\textwidth}   
            \centering 
            \includegraphics[scale=0.3]{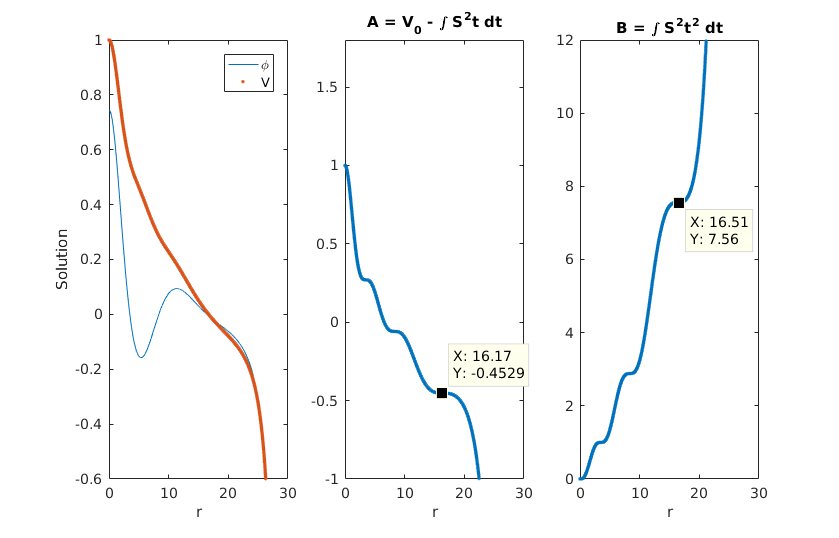}
            \caption[]%
            {{\scriptsize $S_0 =0.74428221769954627796 $, $A = -0.4529 $, $B = 7.56$}}    
            \label{fig:mean and std of net34}
        \end{subfigure}
        \quad
        \begin{subfigure}[b]{0.475\textwidth}   
            \centering 
            \includegraphics[scale=0.3]{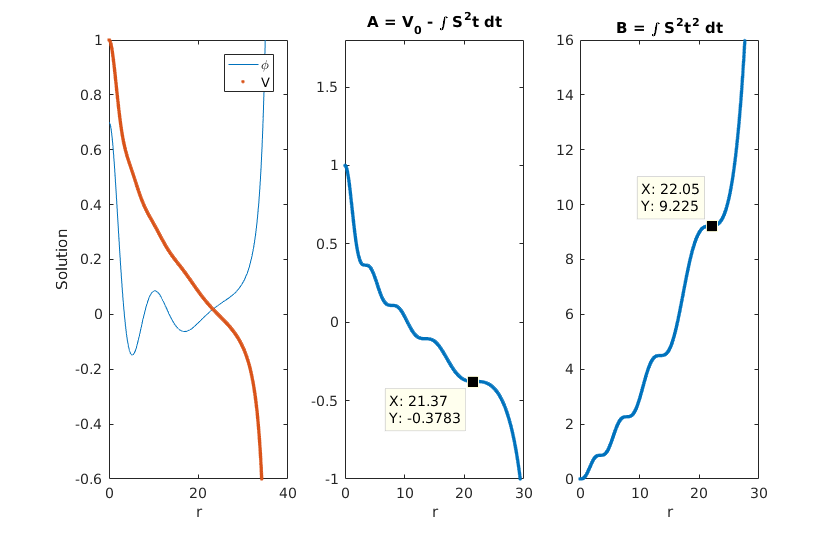}
            \caption[]%
            {{\scriptsize $S_0 =0.70000260975984696099 $, $A =-0.3783 $, $B =9.225 $ }}    
            \label{fig:mean and std of net44}
        \end{subfigure}
        \caption[   ]
        {\small The first four bound-state wave functions for deformed 
        Hamiltonian with $\beta = -10^{66}$ and the $A$, $B$ value for each Wavefunction. } 
        \label{beta=-1e66}
    \end{figure*}

\begin{figure*}
        \centering
        \begin{subfigure}[b]{0.475\textwidth}
            \centering
            \includegraphics[scale=0.3]{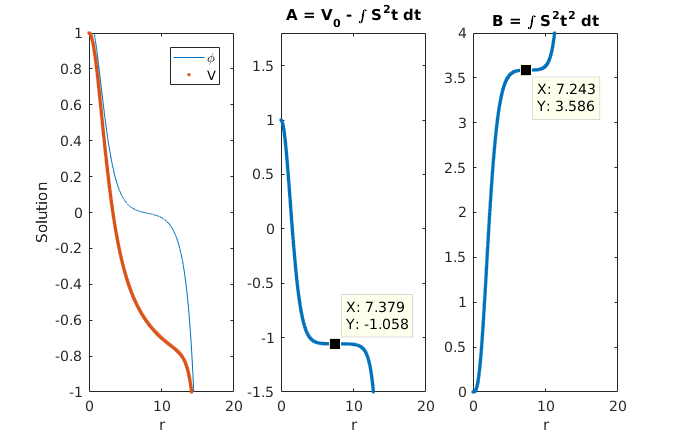}
            \caption[]%
            {{\scriptsize $S_0 =1.0887152281332919124 $, $A = -1.058 $, $B = 3.586$}}    
            \label{fig:mean and std of net14}
        \end{subfigure}
        \hfill
        \begin{subfigure}[b]{0.475\textwidth}  
            \centering 
            \includegraphics[scale=0.3]{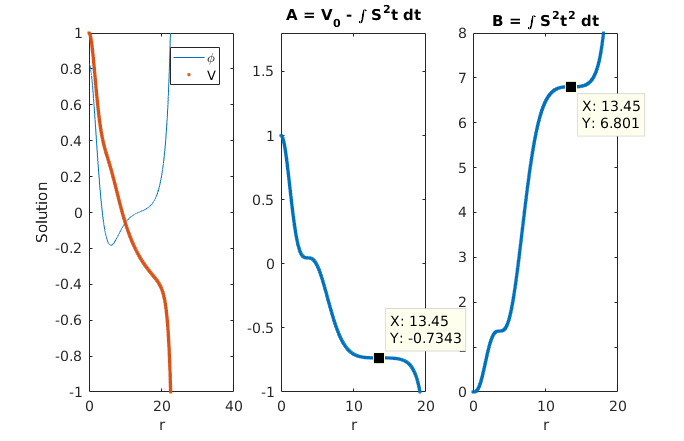}
            \caption[]%
            {{\scriptsize $S_0 =0.82649368253245469873$, $A = -0.7343 $, $B = 6.801$}}    
            \label{fig:mean and std of net24}
        \end{subfigure}
        \vskip\baselineskip
        \begin{subfigure}[b]{0.475\textwidth}   
            \centering 
            \includegraphics[scale=0.3]{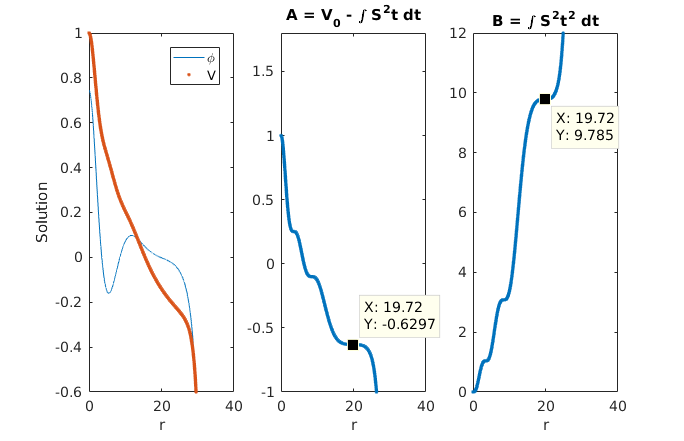}
            \caption[]%
            {{\scriptsize $S_0 =0.74422065844062346152$, $A = -0.6297 $, $B = 9.785$}}    
            \label{fig:mean and std of net34}
        \end{subfigure}
        \quad
        \begin{subfigure}[b]{0.475\textwidth}   
            \centering 
            \includegraphics[scale=0.3]{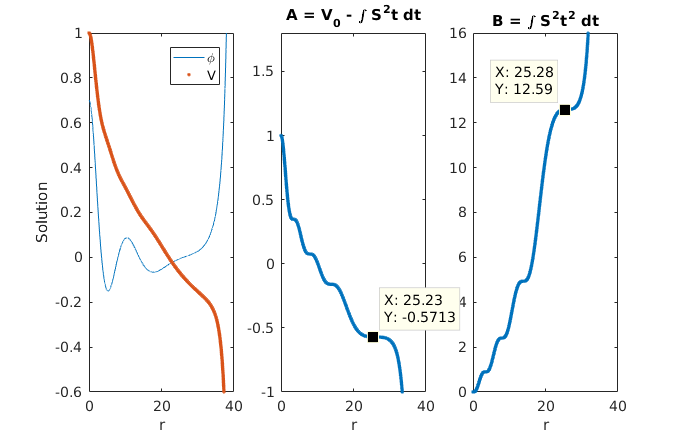}
            \caption[]%
            {{\scriptsize $S_0 =0.70014674500844686911$, $A =-0.5713 $, $B =12.59 $ }}    
            \label{fig:mean and std of net44}
        \end{subfigure}
        \caption[   ]
        {\small The first four bound-state wave functions for deformed 
        Hamiltonian with $\beta = -10^{65}$ and the $A$, $B$ value for each wave function.} 
        \label{beta=-1e65}
    \end{figure*}

The numerical results for $\beta = -10^{66}$ and $\hbar^2\beta=-1.112\times 10^{-2}$
and those for $\beta = -10^{65}$ and $\hbar^2\beta=-1.112\times 10^{-3}$ are shown 
in Figure \ref{beta=-1e66} and Figure \ref{beta=-1e65}, respectively. Here, we showed
the first four bound-states for each $\beta$. The fist plot in each subfigure shows
the solutions for $S$ (blue line) and $V$ (red line).  The second plot in each 
subfigure has the meaning of the infinite limit of $V$, named $A$; and the third
plot is related to the rescaling factor $\lambda$.
Following the same standard as that in \cite{moroz_spherically}, we fix the the 
bound-states which mark transitions between solutions in which $(S, V)\rightarrow (+\infty, -\infty)$ and those in which $(S, V)\rightarrow (-\infty, -\infty)$. The normalization requirement is that the summation of the probability in the whole space is equal to $1$. Therefore, one should have
\begin{equation}
1 =  \int^{\infty}_0 4\pi r^2 |\psi|^2 d r = \int^{\infty}_0 4\pi (\lambda r)^2  \left| \frac{\zeta S}{\lambda^2}\right|^2 d (\lambda r) = \frac{4\pi\zeta^2 B }{\lambda},
\end{equation}
where $B = \int^{\infty}_0 r^2  \left| S\right|^2 d r $. Therefore, the rescaling
parameter should be $\lambda =4\pi\zeta^2 B $. To see the limit of $V$ at the 
infinity, $V$ can be expanded in powers of $r^{-1}$ (only to the first order):
\begin{equation}
V = A + \frac{B}{r} + \cdots,
\end{equation}
where
\begin{align}
A = V_0 - \int^{\infty}_0 r S^2 d r.
\end{align}
Therefore, the limit of $V$ at infinity is $A$. Since at the infinity, 
the potential energy is approximate zero, we can determine the energy
eigenvalues by normalizing the $A$ and also multiplying the coefficient $\zeta$.
Therefore, we can obtain the 
\begin{equation}
E = \xi \frac{A}{\lambda^2} = \frac{\xi A}{\zeta^4 (4\pi B)^2}.
\end{equation}


Even though we can obtain the numerical result for $\beta <0$, however, 
there are two obstacles preventing us to obtain good results.
The first one is the error of the initial value for the
second order derivative of $S_0$. Because of the existence of the deformed term, 
the highest order of the derivative of $S$ is the fourth order. Using fourth-order Runge-Kutta
NAG routine, we need to give the initial values for the derivatives. Since $C_{2i+1}=0$,
we know the first order and third order derivatives of the $S_0$ are both zero.
The coefficient $C_2$ is equal to $-S_0/6$ plus corrections from the non-zero 
$\beta C_4$ (see \eqref{coefficient}) and hence the  second derivative 
will be $-S_0/3$ plus corrections by some factor proportional to $\hbar^2\beta$ 
(leading term). However, one can not write down the exact correction for $S_0^{2}$, 
since to solve the exact $C_4$, we need to know $C_6, C_8$ and so on. Therefore, when we 
set the second derivatives $S_0$ to be $-S_0/3$, there is an initial error for 
$S_0^{(2)}$ which is proportional to $\hbar^2\beta$. The second obstacle
is the error from the iterations. When we perform the numerical calculation, the 
coefficient $2\hbar^2\beta$ in front of $S^{(4)}$ can be treated as $1$ and meanwhile 
we multiply a coefficient $\frac{1}{2\hbar^2\beta}$ to the lower orders. 
Therefore, we know that by performing iterations, there is a error proportional 
to  $\frac{1}{2\hbar^2\beta}$. Therefore, if we choose a small $\beta$, the error
from the second obstacle would be significant; if we choose a big $\beta$, 
the error from the first one would be not ignorable. Moreover, as we before said, 
for $\beta \geq 0$, we even can not find a stable solution. Thus, there are various 
problems with the application of the usual Runge-Kutta method to this system. 

\section{Two-Step Runge-Kutta Method}
In the previous  we observe that the usual Runge-Kutta method could not be used 
to analyze this system. So, in this section, we will develop a new method, which we 
shall call a two-step Runge-Kutta method, and it will be demonstrated that 
this new two-step Runge-Kutta method can be used to analyze this system. 
there are various problems with the application of the usual Runge-Kutta method. 
The main idea behind this two-step Runge-Kutta method is to perform the numerical analysis 
in two steps. In the first step the 
usual  the numerical method are used to  solve the un-deformed theory,  which can
directly remove the error form the two obstacles  discussed in the previous 
section. This is followed by a second step in which the numerical 
solutions obtained for the un-deformed 
theory are deformed by a perturbation generated from the GUP deformation of the original 
theory. 
We call this approach which is based on two steps, as a two-step Runge-Kutta method.

So, for a undeformed theory, we have 
\begin{align}\label{dimensionless_emo_un}
\frac{d^2}{dr^2}(rS)  = - rVS,\\
\frac{d^2}{dr^2}(rV) = -rS^2.
\end{align}
The formulas \eqref{dimensionless_emo_un} is invariant under a scale transformation:
\begin{equation}\label{rescaling_un}
(S, V,  r) \rightarrow (\lambda^2 S, \lambda^2 V, \lambda^{-1}r).
\end{equation}
Now because of the rescaling freedom \eqref{rescaling_un}, one  can also fix $V_0 =1$ and 
allow $S_0$ to vary. Then, following a similar derivation as that \eqref{coefficient} 
in the deformed case, one can show that the power-series expansions for $S$ and $V$ are:
\begin{align}\label{power-series}
\begin{aligned}
& S = S_0 -\frac{1}{6}S_0 r^2 + \frac{1}{120}S_0(S_0^2 +1)r^4 + \cdots, \\
& V = 1 -\frac{1}{6} + \frac{1}{60}S^2_0 r^4 +\cdots.
\end{aligned}
\end{align}
Therefore, the first derivatives of $S$ and $V$ at $r=0$ are both zero. 
Using a standard fourth-order Runge-Kutta NAG routine, we can resolve 
the formulas \cite{moroz_spherically}. 
\begin{figure*}
        \centering
        \begin{subfigure}[b]{0.475\textwidth}
            \centering
            \includegraphics[scale=0.3]{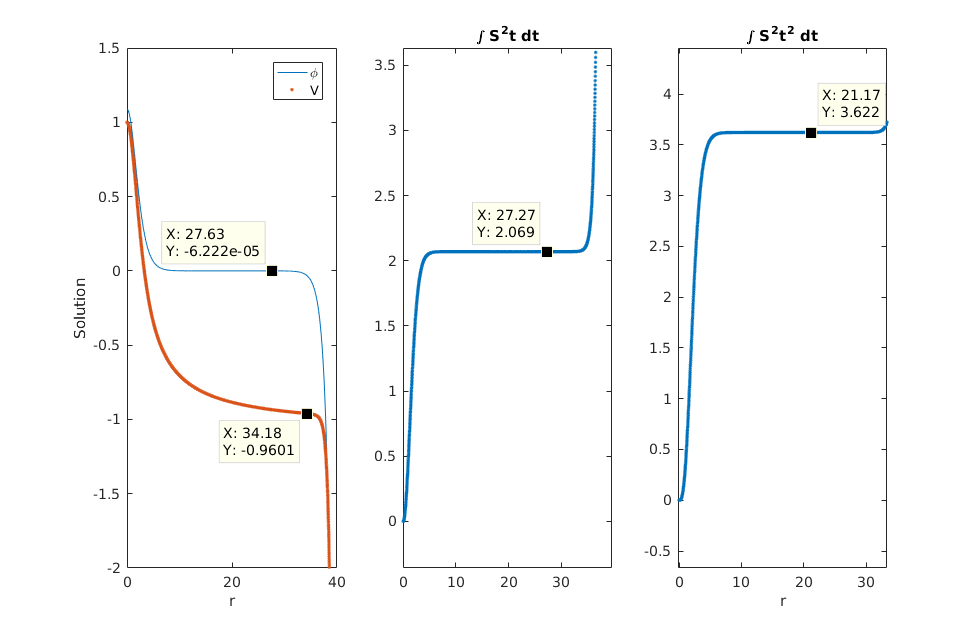}
            \caption[]%
            {{\scriptsize $S_0 =1.0886370794286974739 $, $A =V_0 -\int^{\infty}_0 r S^2 d r = -1.069 $, $B = 3.622$}}    
            \label{fig:mean and std of net14}
        \end{subfigure}
        \hfill
        \begin{subfigure}[b]{0.475\textwidth}  
            \centering 
            \includegraphics[scale=0.3]{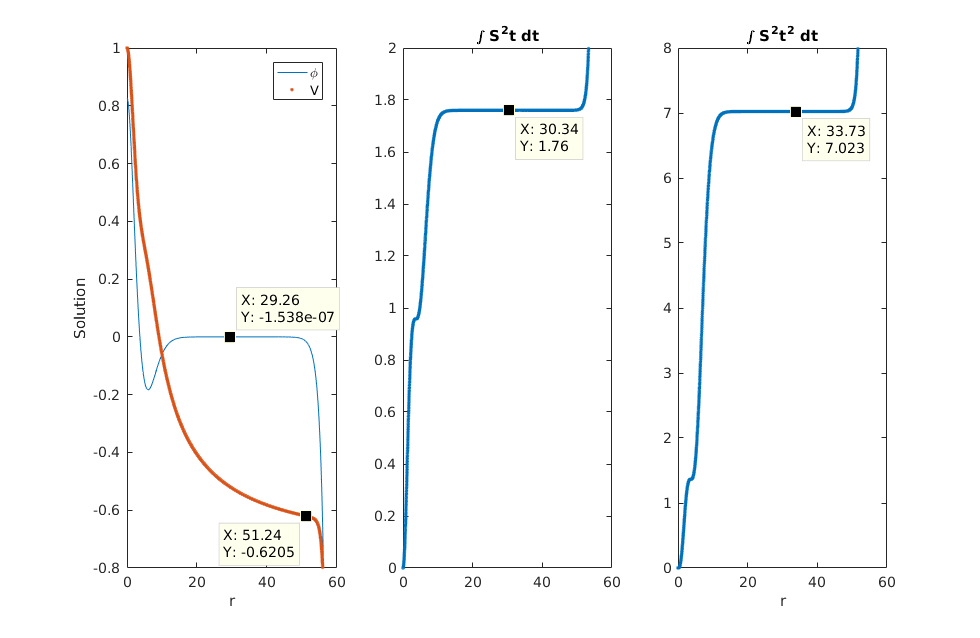}
            \caption[]%
            {{\scriptsize $S_0 =0.82647428414036172573 $, $A = -0.76 $, $B = 7.023$}}    
            \label{fig:mean and std of net24}
        \end{subfigure}
        \vskip\baselineskip
        \begin{subfigure}[b]{0.475\textwidth}   
            \centering 
            \includegraphics[scale=0.3]{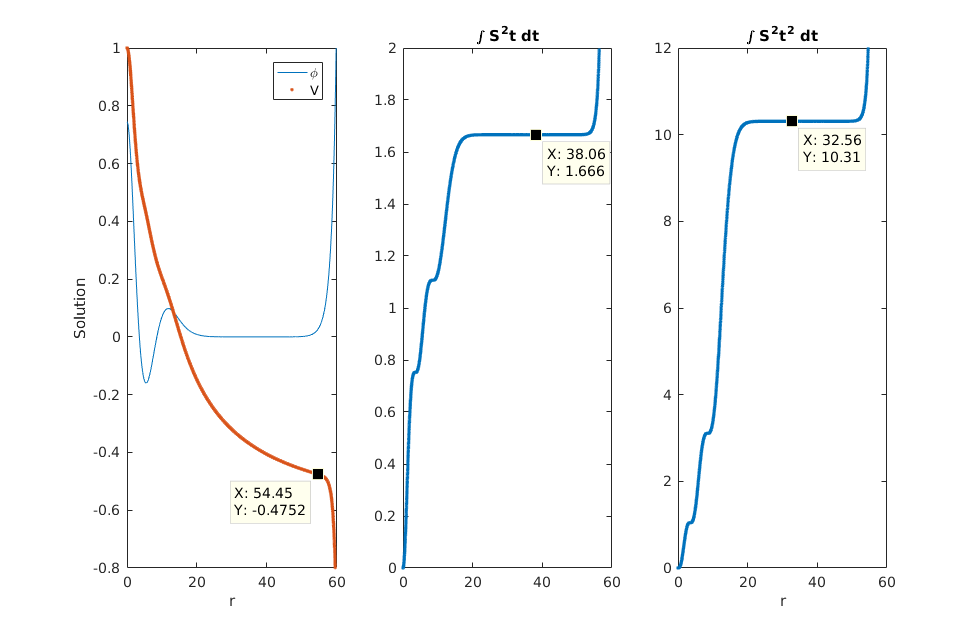}
            \caption[]%
            {{\scriptsize $S_0 =0.74421337845029078562 $, $A = -0.666 $, $B = 10.31$}}    
            \label{fig:mean and std of net34}
        \end{subfigure}
        \quad
        \begin{subfigure}[b]{0.475\textwidth}   
            \centering 
            \includegraphics[scale=0.3]{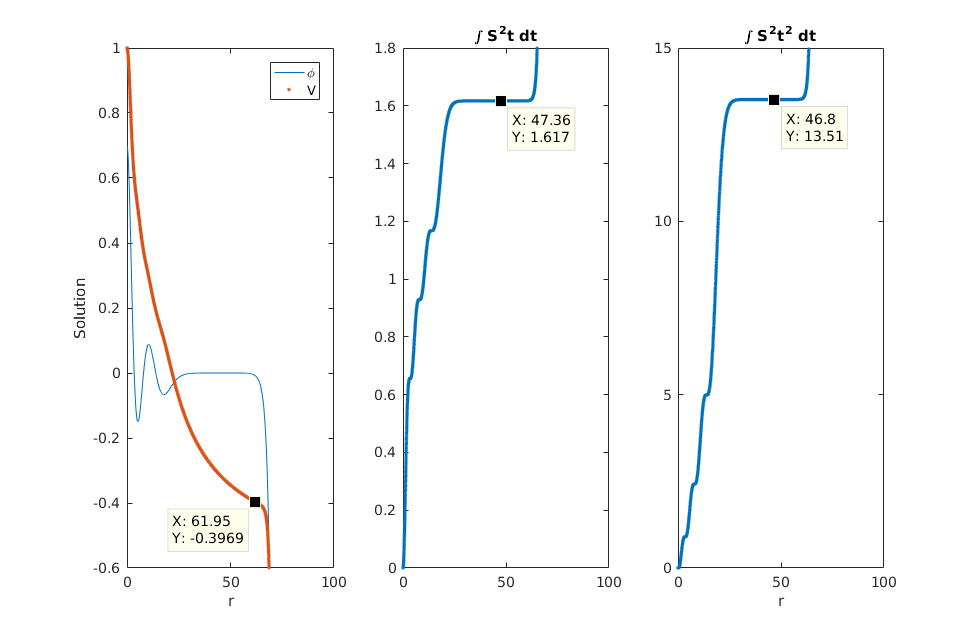}
            \caption[]%
            {{\scriptsize $S_0 =0.70014479719102051813 $, $A =-0.617 $, $B =13.51 $ }}    
            \label{fig:mean and std of net44}
        \end{subfigure}
        \caption[   ]
        {\small The first four bound-state wave functions.
        $S_0$ is the approximate amplitude of the Wavefunction
        at $r=0$. $A$ is approximate value of $V$ at infinity. 
        $B$ is related to the normalization factor. } 
        \label{beta=0}
    \end{figure*}
 See figure \ref{beta=0} and we showed the solutions for the first 
 four wave functions. The solutions for $S$ and $V$ are shown in the 
 first plot of each subfigure; the limit $A$ of $V$ at infinity and 
 $B$ are shown in the second and the third plot of each subfigure, respectively. 
 Then we can determine the energy eigenvalues by normalizing the $A$ and also 
 multiplying the coefficient $\zeta$. Therefore, we obtain
\begin{equation}
E = \xi \frac{A}{\lambda^2} = \frac{\xi A}{\zeta^4 (4\pi B)^2}.
\end{equation}

Now, let us consider the  correction due to 
the perturbation term $H_1$. Since  
\begin{align}
H_1 =  \frac{ \beta}{ m} \hat{p}^4, && H_0 = \frac{\hat{p}^2}{2 m} +\Phi,
\end{align} 
we have
\begin{equation}
H_1 = (4\beta m)\left[H_0^2 + \Phi^2 - (H_0 \Phi + \Phi H_0)\right].
\end{equation}
At the infinity, the potential energy goes to zero, for specific eigenstates, we have 
\begin{equation}
\Delta E_{0n} = \delta\left\langle\psi_n |H_1|\psi_n \right\rangle = 4\beta m E_n^2.
\end{equation}
Considering the numerical result we just obtained, we know that 
\begin{equation}\label{correction_theoritic}
\frac{\Delta E_{0n}}{E_{0n}} = 4\beta m E_n = 4m (\lambda^2 \beta)
\left(\xi \frac{A_{0n}}{\lambda^2}\right) = 2\hbar^2\beta A_{0n}. 
\end{equation}
Let's call this relative difference of the eigenvalue the two-step numerical 
difference. Since we also directly computed the numerical values of the eigenvalues, 
we can compare the numerical difference with the two-step numerical difference. 
The numerical difference is 
\begin{align}
\begin{aligned}
\frac{\Delta E_{0n}}{E_{0n}} = \frac{E_n - E_{0n}}{E_{0n}} = \frac{\frac{A_n}{B^2_n}- \frac{A_{0n}}{B^2_{0n}}}{\frac{A_{0n}}{B^2_{0n}}}. 
\end{aligned}
\end{align}
The comparison for $\beta=-10^{66}$ is shown in table \ref{table_beta=-1e66}. 
The comparison for $\beta=-10^{65}$ is shown in table \ref{table_beta=-1e65}.
\begin{table}\caption{$\beta=-10^{66}$.}
\begin{center}
\begin{tabular}{ |c|c|c|c| } 
\hline
& $\frac{\Delta E_{0n}}{E_{0n}}$ (numerical) & $\frac{\Delta E_{0n}}{E_{0n}}$ 
(semi-theoretic) & difference \\
\hline
1 & 0.07437 & 0.02378 & $212\%$ \\ 
2 & 0.18585 & 0.01690 & $1000\%$\\ 
3 & 0.26474 & 0.01481 & $1687\%$ \\ 
4 & 0.31501 & 0.01372 & $2195\%$ \\ 
\hline
\end{tabular}
\end{center}
\label{table_beta=-1e66}
\end{table}
\begin{table}\caption{$\beta=-10^{65}$.}
\begin{center}
\begin{tabular}{ |c|c|c|c| } 
\hline
& $\frac{\Delta E_{0n}}{E_{0n}}$ (numerical) & $\frac{\Delta E_{0n}}{E_{0n}}$
(semi-theoretic) & difference \\
\hline
1 & 0.07437 & 0.002378 & $307\%$ \\ 
2 & 0.18585 & 0.001690 & $1692\%$\\ 
3 & 0.26474 & 0.001481 & $3254\%$ \\ 
4 & 0.31501 & 0.001372 & $4725\%$ \\ 
\hline
\end{tabular}
\end{center}
\label{table_beta=-1e65}
\end{table}
Considering the tables \ref{table_beta=-1e66} and \ref{table_beta=-1e65},
we can see that for the first modes, the difference between the numerical 
$\frac{\Delta E_{01}}{E_{01}}$ and its two-step numerical value are quite small 
than other modes. At least, the numerical value and the theoretic value are 
in the same order of the magnitude. For the other modes, the numerical results are quite bad. 
The higher level of mode, the bigger difference. This might be because of the reason 
that for the higher modes, the initially error for $S^{(2)}_0$ would be amplified larger.
Moreover, if we restrict $\beta$ to satisfy the bound from the experiment, $\beta$ would 
be positive and the absolute value of $\beta$ should be even small than what we chose. 
For this case, the numerical approach can not give us a solution. However, the two-step numerical 
approach can be suitable to general $\beta$. 
Therefore, we see that the two-step numerical approach has more applications.
Thus, the two-step numerical approach resolves the problems with the usual 
Runge-Kutta method.
\section{Conclusion}
It is known that because of T-duality, the spacetime geometry in  string theory has an intrinsic minimal length associated with it. 
So, it  is expected that the short distance behavior of general relativity should also be modified in such a way that there 
is an intrinsic minimal length associated with it.  It is also possible to have an optical analog for such short distance modification. 
So, in this paper,   we have  analyzed a short distance  deformation of a  semi-classical 
gravitational system with an intrinsic minimal length. In this system the 
 gravitational field was  treated  as 
a classical field, and it was sourced  by quantum mechanical 
matter fields. As such a system would be described by 
the   Schrodinger-Newton equation, we analyzed its short distance modification 
by analyzing the GUP deformation of the 
   Schrodinger-Newton equation. As the optical analog of Schrodinger-Newton equation means that such a system 
   can be studied in laboratory using its optical analog. It was observed  that the usual 
fourth-order Runge-Kutta method did not work for such a system. This motivated us 
to propose a 
new   two-step Runge-Kutta method   
for analyzing this system.  
In this two-step Runge-Kutta method, the numerical analysis was  perform 
in two steps. In the first step,  the 
usual  the numerical method were used to obtain the solution for 
the un-deformed theory.  
This is followed by a second step, and in that second step,  the numerical 
solutions obtained for the un-deformed 
theory were deformed by a perturbation. This perturbation was 
generated from the GUP deformation of the original 
theory. It was observed that this two-step Runge-Kutta method resolved the problems 
associated with the one step Runge-Kutta method.

This method can be used for studying other similar physical systems.
It is expected that the GUP deformation of  Schrodinger  equation with any  potential 
will have the same problems associated the GUP-deformation of the 
 Schrodinger-Newton equation. Thus, it would not be possible to use the usual 
 Runge-Kutta method for analyzing such a system. However, the two-step Runge-Kutta method, 
 we proposed in this paper can be easily used for analyzing such a system.  
 It may be noted that it is possible to consider a different from of the deformation 
 of the uncertainty principle \cite{faiz}.  This deformation of the uncertainty principle 
  produces a linear term derivative in the Schrodinger equation. It would be interesting 
  to perform such a deformation of the  Schrodinger-Newton equation and analyze the 
  consequences of such a deformation. It would also be interesting to analyze if the usual 
  Runge-Kutta method  or the two-step Runge-Kutta method  proposed in this paper, can 
  be used to analyze such a deformation of the uncertainty principle. 
  It would also be interesting to find    an optical analog for 
 such a deformation.
 
 \section*{Acknowledgements}
 The work of Q.Z. is supported by NUS Tier 1 FRC Grant R-144-000-360-112.

\end{document}